# A Hybrid Similarity-Aware Graph Neural Network with Transformer for Node Classification

Aman Singh[†,a], Shahid Shafi Dar[†,a], Ranveer Singh[a] and Nagendra Kumar[a,*]

[a]Department of Computer Science and Engineering, Indian Institute of Technology Indore, 453552, India

## Abstract

Node classification has gained significant importance in graph deep learning with real-world applications such as recommendation systems, drug discovery, and citation networks. Graph Convolutional Networks and Graph Transformers have achieved superior performance in node classification tasks. However, the key concern with Graph Convolutional Networks is over-squashing, which limits their ability to capture long-range dependencies in the network. Additionally, Graph Transformers face scalability challenges, making it difficult to process large graphs efficiently. To address this, we propose a novel framework, A Hybrid **SI**milarity-Aware **G**raph Neural **Net**work with Transformer for Node Classification (SIGNNet), which capitalizes local and global structural information, enhances the model's capability to effectively capture fine-grained relationships and broader contextual patterns within the graph structure. The proposed method leverages Graph Convolutional Networks alongside a score-based mechanism to effectively capture local and global node interactions while addressing the limitations of over-squashing. Our proposed method employs a novel Personalized PageRank-based node sampling method to address scalability issues by generating subgraphs of nodes. Additionally, SIGNNet incorporates a novel attention mechanism, Structure-Aware Multi-Head Attention (SA-MHA) which integrates node structural information for informed attention weighting, enabling the model to prioritize nodes based on topological significance. Extensive experiments demonstrate the significant improvements achieved by the proposed method over existing state-of-the-art methods, with average accuracy gains of 6.03%, 5.47%, 4.78%, 19.10%, 19.61%, 7.22%, 19.54% and 14.94% on Cora, Citeseer, CS, Wisconsin, Texas, Actor, Cornell and Chameleon datasets, respectively.

## 1. Introduction

Graphs are essential tools for modeling complex, non-euclidean relationships between entities, making them highly valuable in various real-world applications [1]. These relationships in the graphs can be categorized into homogeneous and heterogeneous types. In homogeneous graphs, nodes of the same type are connected to each other whereas, in heterogeneous graphs, nodes of different types may be connected. Graph-based representations are important in tasks such as social network analysis, recommendation systems, biological network analysis, and much more [2]. Such representations enable improved relational data modeling and provide insights into community detection, node classification, and edge prediction. The ability to represent entities and their interactions in graph form makes them indispensable for capturing the inherent relational structure in diverse fields.

Node classification involves predicting the labels of nodes within a graph by leveraging both their features and the network topology. Our objective is to effectively classify nodes in both homogeneous and heterogeneous graphs, addressing the unique challenges inherent to each graph structure. In homogeneous graphs, the challenge primarily lies in learning from uniform node and edge types, while heterogeneous graphs require capturing complex, multi-typed relationships and node attributes. To address these challenges, Graph Neural Networks (GNNs) have emerged as robust and effective models for leveraging the structural and feature-based information in graph data. GNNs excel in node classification tasks by capturing both local and global patterns within the network, making them particularly well-suited for handling the complexities of the graph structures. Among GNNs, Graph Convolutional Networks (GCNs) [3] have gained significant popularity. GCNs propagate messages between neighboring nodes, aggregating features into unified representations. This localized feature aggregation enhances the expressiveness of node embeddings, enabling better performance in node classification [4], community detection [5], recommendation systems [6], and

---

graph classification [7] tasks. Despite their success, GCNs perform optimally on homophilic graphs, where nodes share high feature similarity with their neighbors. However, they face challenges when applied to heterophilic graphs, where neighboring nodes differ significantly in their attributes. This limitation, known as over-squashing [8], occurs when the model struggles to propagate information across distant nodes, thus failing to capture long-range dependencies critical for effectively modeling complex, heterogeneous relationships in graphs. The Graph Transformer was introduced to overcome the limitations of GCNs, demonstrating superior performance in node classification [9]. Graph Transformer captures information from the entire graph, allowing nodes to aggregate knowledge from all other nodes through an attention mechanism. This global attention is advantageous for node classification, as it helps in capturing complex dependencies across the graph. However, because each node computes attention scores with every other node, the approach becomes computationally expensive and may sometimes capture irrelevant information. This inefficiency can reduce node classification effectiveness, as not all nodes contribute meaningfully to the representation of a target node. To address this, our proposed framework selectively focuses on the most relevant nodes, enhancing classification performance and computational efficiency.

Existing methods such as SAN [10] and SNGNN++ [4], offer state-of-the-art performance by addressing the common limitations of GCNs and transformers, ensuring accurate and contextually relevant outcomes. These methods use raw node features that contain only attribute information and lack local and global structural context. This limitation impairs the model's capacity to fully capture the underlying graph topology and the intricate relationships between nodes. Additionally, computing attention over the entire graph introduces scalability challenges due to the high computational cost of processing all node pairs, making these methods inefficient for large graphs. Another limitation is that these methods calculate attention scores based solely on node feature similarity, neglecting critical structural information. Attempts to capture the network using positional encodings, such as laplacian and spatial encodings, often fail to fully capture the graph structure. This restriction hampers the capture of crucial relationships between distant nodes, leading to less effective results.

To address the challenges mentioned above, we introduce a novel framework, SIGNNet, a Similarity-Aware Graph Neural Network for node classification. Our proposed method capitalizes GCNs to aggregate local structural properties into node representations, enhancing node classification by refining features and enabling the model to better distinguish between classes based on local patterns. However, GCNs' dependency on local neighborhood information limits their ability to capture long-range dependencies, which are essential for fully understanding the graph structure. To address the limitations of local dependency in GCNs, we employ a compatibility matrix that evaluates how similar one node is to another across the entire graph. This matrix helps the model capture a more comprehensive network context, enhancing its ability to understand long-range dependencies crucial for a complete analysis of the graph's structure. Furthermore, we utilize class-centric features that simplify and highlight the essential characteristics of each class, allowing the model to more effectively distinguish nodes based on these unique patterns. In addition, we incorporate a connection score, a crucial measure of node importance, to further refine the node features. This integration enhances the ability of model to detect semantic and structure relationships and emphasize node significance, thereby improving the classification performance. To tackle scalability, we propose a novel sampling technique based on the Personalized PageRank (PPR) algorithm [11]. This method constructs node representations by leveraging features from a sampled subgraph, rather than processing the entire graph. This approach mitigates scalability challenges by reducing the input size from $|V|$ (number of nodes in the dataset) to $|K|$ (the average number of nodes in a subgraph), where $|V| >> |K|$. However, the use of subgraphs may lead to a partial loss of global structural information. To address this, we introduce Structure-Aware Multi-Head Attention (SA-MHA), which enhances the transformer's attention mechanism by incorporating both node feature similarity and the graph's structural properties. This ensures a more accurate representation of the overall network. Overall, by combining node sampling with feature enhancement, our approach solves the scalability problem and improves node classification performance graph. Our results demonstrate that SIGNNet consistently outperforms existing methods. The main contributions of SIGNNet are as follows:

1. We present a novel framework that capitalizes on the structural and semantic details of the graph to enhance node features, enabling a more significant understanding of node interactions and resulting in improved model performance.

2. We propose a class-centric approach in conjunction with a connection score-based method to capture higher-order contextual information across the network, leading to globally enriched node feature representation.

3. We propose a Structure-Aware Multi-Head Attention (SA-MHA) that integrates network structural information directly into the attention computational process, improving the representation of complex dependencies within the network.

4. We propose a node-sampling technique based on the Personalized PageRank algorithm to generate localized subgraphs for each node, enabling efficient scalability during training while preserving essential structural properties of the graph.

5. We introduce neighborhood-influenced feature learning to aggregate spatial dependencies, facilitating more robust node feature representation. This approach improves the model's capacity for capturing localized structural patterns.

6. To thoroughly assess the performance of our proposed method, we have carried out extensive experiments across eleven benchmark datasets. The results indicate that our approach consistently achieves better outcomes than the current state-of-the-art methods.

The overall flow of the paper is as follows: Section 2 presents the literature survey, while the problem statement has been defined in Section 3. The architectural overview and methodology are defined in Section 4. Experimental results and ablation studies are discussed in Section 5. Section 6 discusses the limitations and future direction. Finally, Section 7 summarizes our study.

## 2. Related Work

The rapid growth of graph-based data across different fields, such as social networks [12, 13], and biological networks [14, 15], has significantly advanced the node classification task. This task focuses on predicting labels for nodes by utilizing the attribute value of the nodes and their connections in the network. The complexity and scale of graphs present challenges that have prompted researchers to increasingly focus on deep learning approaches, particularly Graph Convolutional Networks (GCNs) [3], which have shown strong performance in capturing local node dependencies. More recent advancements include the transformer-based approach that leverages attention mechanisms to further enhance the understanding of node relationships in large and complex graphs. We have categorized this study into two general categories: GCNs-based methods and Graph Transformer-based methods.

### 2.1. GCNs-based Methods

GCN is a deep learning framework developed to transform complex graph structures into meaningful representations. It updates node features by aggregating information from neighboring nodes along with their features as shown in Figure 1. The inputs to the Graph Convolutional Network are an adjacency matrix that provides the information how nodes are connected in the network and a feature matrix that provides the attribute information of the nodes. The GCN processes these inputs to produce embeddings for each node. Node embedding contains information on the neighboring node's features and is suitable for node classification [16]. GCN operation can be described in layers, where each layer updates the features of the nodes. Based on the connection defined in the adjacency matrix each node shares its features with its neighbors. Then, in the next layer, it aggregates all the features, usually through a linear transformation followed by a non-linear activation function. This process helps to gather the broader contextual information of the network, and it is repeated over multiple layers. In the end, the GCN generates final node embeddings that provide richer representations for better node classification.

GCNs have significantly advanced node classification tasks on graphs. The APPNP [17] employs a Personalized PageRank-based propagation scheme to enhance node classification by expanding the utilized neighborhood, thereby improving model efficiency. However, while APPNP leverages the Personalized PageRank (PPR) matrix for message passing, it can occasionally introduce redundant propagation, which may result in inefficiencies when capturing finer relationships within complex networks. GCNII [18] enhances traditional GCNs by integrating residual and identity mappings to tackle the over-smoothing problem, allowing for better preservation of feature information across multiple layers. While these improvements help maintain performance, GCNII can still encounter challenges in retaining initial node information, particularly in deeper layers and denser graphs, which may limit its ability to capture complex relationships effectively. Finally, GATE-GNN [19] employs a network ensemble to address imbalanced data, enhancing classification accuracy and stability in challenging environments. While this ensemble-based approach effectively handles class imbalance, the added complexity can reduce its efficiency on large-scale graphs, posing scalability and computational overhead challenges. Each method contributes distinctively to addressing the complexities of graph-structured data in supervised learning contexts. This research seeks to utilize deep learning approaches to create a sturdy and adaptable framework for node classification across diverse scenarios, addressing the complexities of graph-structured data. In the subsequent sections, we will discuss the Graph Transformer-based methods.

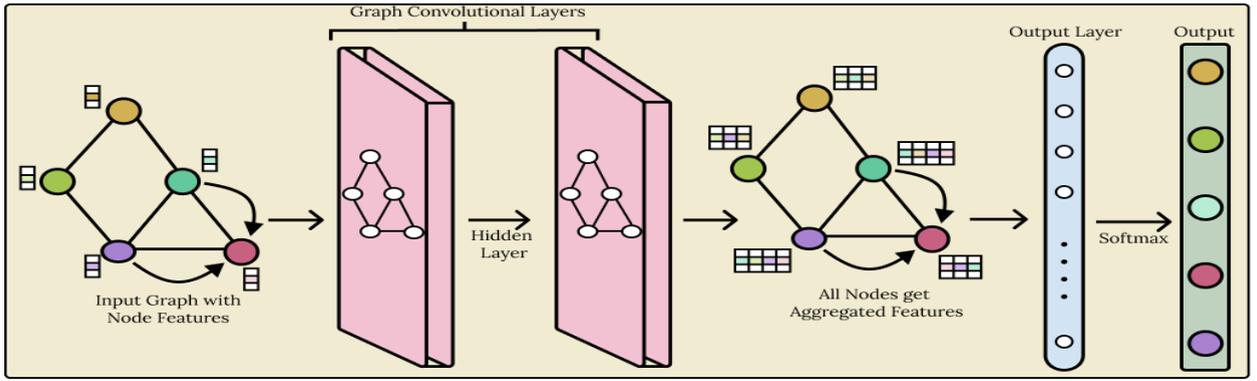

**Figure 1:** This figure shows node classification using Graph Convolutional Networks. It represents how each node shares its feature with the neighboring nodes and gets the aggregated feature. These features are processed and classified in the output layer using a softmax function, resulting in the final classifications.

### 2.2. Graph Transformer-based Methods

Graph Transformer is a powerful and effective model for graph-based tasks, such as edge prediction [20], community detection [21, 22], and node classification, showing superior performance in these applications. The core components of a Graph Transformer are the Multi-Head Attention (MHA) mechanism and a position-wise feed-forward network. The MHA mechanism operates on an input sequence of node features, where each node's feature is projected into query, key, and value spaces using learnable weight matrices. This enables the model to apply attention by computing interactions between the projected features, thereby capturing diverse node relationships. By processing multiple attention heads simultaneously, the MHA mechanism captures a broader range of contextual information within the graph [23, 24]. The outputs from various attention heads are concatenated and undergo a linear transformation to create a refined embedding of the node features. This design allows Graph Transformers to effectively learn complex patterns and dependencies in graph structures, making them highly effective for node classification.

Graph Transformer-based methods for node classification have introduced innovative ways to enhance node representation and scalability. Gapformer [25] integrates Graph Transformer with graph pooling, significantly reducing computational complexity while maintaining long-range interactions. However, while the pooling mechanism enhances efficiency, it may occasionally overlook critical local node information, which can impact performance in fine-grained node classification tasks. Dual-Encoding Transformer [26] utilizes both structural and semantic encoders to enhance representation learning by capturing comprehensive node interactions. However, the integration of dual encoders can introduce synchronization challenges, especially in highly connected or large-scale graphs where balancing structural and semantic information becomes complex. Spectral Attention Network (SAN) [10] leverages Laplacian-based positional encoding to mitigate over-squashing and enhance sub-structure detection, offering strong global context modeling. However, while utilizing the full Laplacian spectrum improves global representation, this reliance may limit its ability to capture localized structural details, particularly in graphs with distinct local variations.

## 3. Problem Definition

we introduce a novel framework for node classification within graph-based structures. The objective is to accurately classify nodes by extracting, enhancing, and leveraging their inherent features, thereby capturing underlying structural and relational information embedded in the graph structures. Mathematically, a graph is defined as $G = (V, E, X)$, where $V = \{v_1, v_2, \ldots, v_n\}$ represents a finite set of $n$ nodes, $E$ denotes the set of edges capture the relationships between nodes, and $X = [x_1, x_2, \ldots, x_n] \in \mathbb{R}^{n \times d}$ is the node feature matrix, with $d$ denoting the dimensionality of the feature space. The vector $x_i$ is the feature vector of node $v_i$, which represents its characteristics.

The goal is to predict the node's label using a supervised learning framework. This framework will construct a function $f$ that maps input features $X$ to the corresponding node labels $Y$. The set $Y$ includes all possible node labels, represented as $\{y_1, y_2, \ldots, y_m\}$. The framework is designed to improve feature representations, helping it to work well across different nodes and predict their labels correctly. The main goal is to create a mapping that is both accurate and flexible, ensuring better classification results for different graph structures.

This research aims to accomplish the following goals:

- Investigate the use of Structure-Aware Multi-Head Attention to incorporate structural information directly into the attention mechanism, enabling more effective aggregation of network topology in the learning process.

- Examine whether the integration of connection-aware features with class-centric features facilitates the embedding of higher-order network information into node features.

- Whether a node-sampling strategy using the Personalized PageRank algorithm to generate localized subgraphs, aiming to enhance training scalability while retaining essential structural properties of the graph.

- Investigate whether the aggregation of neighborhood-influenced features effectively captures spatial dependencies within node features.

The Table 1 outlines the special symbols used in this paper, providing clarity for the mathematical formulations discussed.

**Table 1**
Overview of symbols

| Notation | Description |
|---|---|
| $G = (V, E, X)$ | Graph |
| $n$ | Number of nodes |
| $d$ | Dimension of node feature vector |
| $V = \{v_1, v_2, \ldots, v_n\}$ | Set of nodes in the graph |
| $X = [x_1, x_2, \ldots, x_n] \in \mathbb{R}^{n \times d}$ | Node feature matrix |
| $P = [p_1, p_2, \ldots, p_n] \in \mathbb{R}^{n \times d}$ | Convolutional feature amplification matrix |
| $X_{\text{final}} \in \mathbb{R}^{n \times d}$ | Final enriched feature matrix |
| $X_{sim} \in \mathbb{R}^{n \times d}$ | Similarity driven feature matrix |
| $X_{deg} \in \mathbb{R}^{n \times d}$ | Connection-based convolutional feature matrix |
| $X_{comp} \in \mathbb{R}^{n \times n}$ | Compatibility matrix |
| $Y = \{y_1, y_2, \ldots, y_m\}$ | Set of node label/class |
| $A \in \mathbb{R}^{n \times n}$ | Adjacency matrix |
| $N(i)$ | Neighbour of node $v_i$ |
| $C(i)$ | Node similar to $v_i$ |
| $W, W_Q, W_V, W_K$ | Learnable weight matrices |
| $A^T$ | Transpose of matrix A |
| $\bar{y}_k$ | Representative of class k |
| $concat$ | Concatenation |

## 4. Methodology

This section focuses on the details of SIGNNet, as shown in Figure 2. The framework has three main parts: (a) feature augmentation and enrichment; (b) adaptive node sampling integrated with sequence module; and (c) attention-based node classification. In the feature augmentation and enrichment module, node features are enhanced by integrating both local and global network information. Local information of the network is incorporated into the node's feature by using neighborhood-influenced feature learning. Global information is captured in two ways: first, through connection score, which measures a node's importance in the network by counting its direct neighbors; and second, by concatenating the most similar class-centric feature to the node's feature. In the adaptive node sampling integrated with the sequence module, we construct node sequences by sampling relevant neighbors to capture the node's structural property and neighborhood information. The final enriched node features and node sequences are passed to the attention-based node classification module, including the Graph Transformer and a dense layer for final prediction.

## 4.1. Feature Augmentation and Enrichment

Traditional node classification methods [27, 28, 29] rely on raw node features, which can overlook complex relationships in graph structures. In the proposed method, node features are enhanced by integrating both local and global network information. Local information is aggregated by leveraging graph convolutional networks while global information is aggregated by connection score and class-centric mechanism, all the steps are defined in Algorithm 1. This makes the feature of each node more comprehensive, resulting in better classification performance. In the next section, we will discuss how local information is incorporated into node features by the message-passing technique.

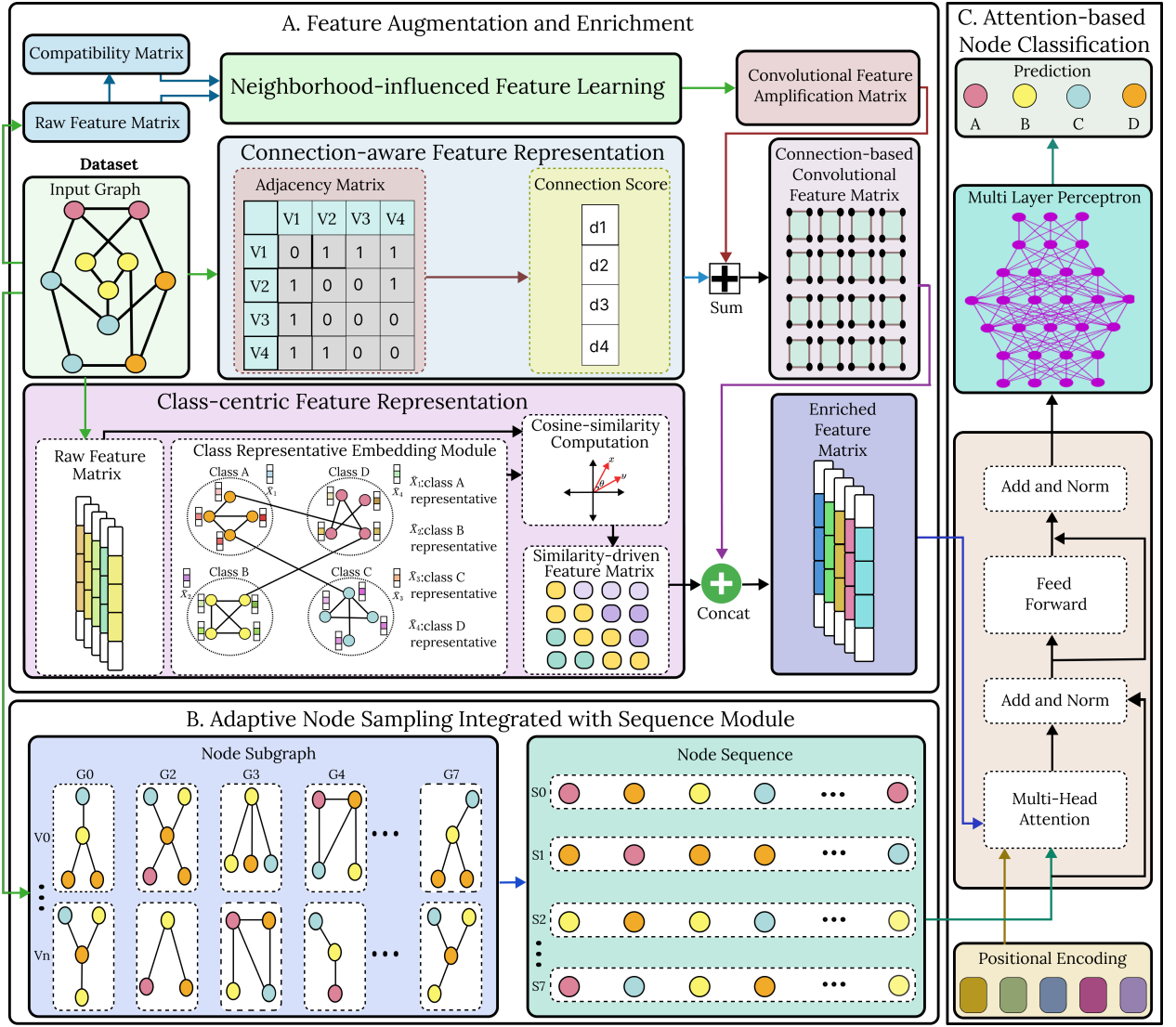

**Figure 2:** Overall architecture of SIGNNet. The process begins with raw features as input in Step (A), where these features are enriched through neighborhood-influenced feature learning, combined with connection scores and class-centric features. In Step (B), adaptive node sampling is applied to generate samples for each node using a sequence module. Finally, the outputs from Steps (A) and (B) are combined in Step (C), which incorporates a graph transformer followed by a neural network for final classification.

### 4.1.1. Neighborhood-influenced Feature Learning

Graph Convolutional Networks (GCNs) is a neural network that works on graphs, capturing the relationship between connected nodes. Each node updates its features by gathering information from its neighbors. The input to the GCNs is feature matrix $X \in \mathbb{R}^{n \times d}$, where $n$ is the number of nodes and $d$ is the dimension of the feature vectors and a compatibility matrix $X_{comp} \in \mathbb{R}^{n \times n}$ encapsulates pairwise feature similarity, marking entries as 1 when similarity exceeds threshold and 0 otherwise. This approach enhances the representation of nodes by covering long-range dependencies and enriching node features with broader contextual information from the network. There are three major steps in the GCN: aggregation, combination, and updating as shown in Figure 3. In the aggregation step, each node $v_i$ aggregates information from its similar nodes $C(i)$ and itself. This information is weighted based on the degree of the node $v_i$ and $v_j$ denoted by $d_i$ and $d_j$ respectively, adjusting how much each node contributes. In the combination step, the aggregated information is processed through a learnable weight matrix $W^{(l)}$, which refines the node features into more meaningful representations. Finally, in the updating step, a non-linear function is applied to update the features for the next layer as shown in Equation 1:

$$h_i^{(l+1)} = \sigma\left(\sum_{v_j \in C(i) \cup \{i\}} \frac{1}{\sqrt{d_i \cdot d_j}} h_j^{(l)} W^{(l)}\right) \qquad (1)$$

where $h_j^{(l)}$ represents the features of node $v_j$ at layer $l$ and $\sigma$ denotes sigmoid activation function. In summary, GCN is applied to the graph structure for the feature enrichment of each node. This leads to a convolutional feature amplification matrix $P = [p_1, p_2, \dots, p_n] \in \mathbb{R}^{n \times d}$ that effectively captures local structural information. In the following section, we will explore the concept of Connection-aware feature representation.

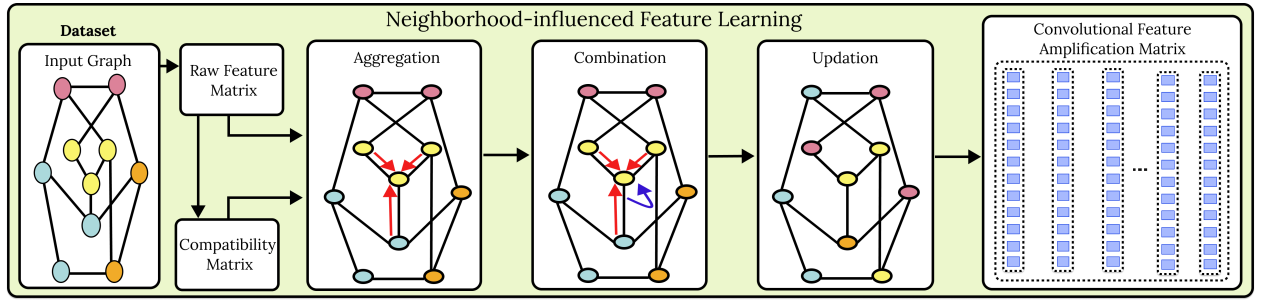

**Figure 3:** Input graph with raw feature and compatibility matrices undergoes neighborhood aggregation, feature combination, and updating, resulting in a convolutional feature amplification matrix for enhanced node representation.

### 4.1.2. Connection-aware Feature Representation

Features are further improved by connection score, a basic measure of a node's importance within the graph. Connection score $deg(v_i)$ evaluates the connectivity of a node by counting the number of adjacent nodes in the network. This metric is very helpful for determining how influential a node is in the network. Mathematically, the connection score for a node $v_i$ is defined in Equation 2.

$$deg(v_i) = \sum_{j=1}^{n} A_{ij} \qquad (2)$$

The adjacency matrix is a square binary matrix of size $A \in \mathbb{R}^{n \times n}$, where $n$ is the total number of nodes in the graph. Each element $A_{ij}$ in the matrix is equal to 1 if there is a direct edge between node $v_i$ and node $v_j$ otherwise, it is 0. This means that $deg(v_i)$ is calculated by summing up all the values in the row corresponding to node $v_i$, effectively counting the total number of edges connected to $v_i$. Including the connection score enriches the node's feature vector by adding information about how well-connected each node is in the graph. This enhancement provides the model with a clearer view of the node's role within the network. Once the connection score is computed for each node, it is integrated into the node's feature vector to form a connection-based convolutional feature matrix $X_{deg} \in \mathbb{R}^{n \times d}$. This integration is performed by adding the connection score $deg(v_i)$ directly to the feature vector $p(v_i)$ of each node $v_i$ obtained by the message passing. The resulting feature vector $x_{deg}(v_i)$ is thus computed as shown in Equation 3.

$$x_{deg}(v_i) = p(v_i) + deg(v_i) \qquad (3)$$

After adding the connection score to each node's feature, the updated node feature now reflects its importance within the network [30], which will contribute to accurate predictions. In the next section, we discuss about class-centric feature representation.

---

**Algorithm 1** Node feature enhancement for graph Transformers

---

**Input:** Adjacency matrix $A \in \mathbb{R}^{n \times n}$, Compatibility matrix $X_{comp} \in \mathbb{R}^{n \times n}$, Feature matrix $X \in \mathbb{R}^{n \times d}$, class labels $\{y_1, y_2, \dots, y_m\}$
**Output:** Final feature matrix $X_{\text{final}} \in \mathbb{R}^{n \times d}$

1: **for** each node $v_i \in V$ **do**
2: $\quad h_i^{(l+1)} = \sigma(\sum_{v_j \in C(i) \cup \{i\}} \frac{1}{\sqrt{d_i \cdot d_j}} h_j^{(l)} W^{(l)})$
3: **end for**
4: $P = [p_1, p_2, \dots, p_n] \in \mathbb{R}^{n \times d}$
5: **for** each node $v_i \in V$ **do**
6: $\quad \deg(v_i) = \sum_{v_j=1}^{n} A_{ij}$
7: $\quad X_{\deg}(v_i) = P(v_i) + \deg(v_i)$
8: **end for**
9: **for** $y_k \in \{y_1, y_2, \dots, y_m\}$ **do**
10: $\quad \bar{x}_k = \frac{1}{|y_k|} \sum_{v_i \in y_k} x_i$
11: **end for**
12: **for** $v_i \in \{v_1, v_2, \dots, v_n\}$ **do**
13: $\quad$ **for** $y_k \in \{y_1, y_2, \dots, y_m\}$ **do**
14: $\quad\quad sim(x_i, \bar{x}_k) = \frac{x_i \cdot \bar{x}_k}{\|x_i\| \|\bar{x}_k\|}$
15: $\quad$ **end for**
16: $\quad k = \arg\max_k sim(x_i, \bar{x}_k)$
17: $\quad X_{sim}(v_i) = [x(v_i), \bar{x}_k]$
18: **end for**
19: $X_{\text{final}} = [X_{\deg}, X_{sim}]$

---

### 4.1.3. Class-centric Feature Representation

We compute class-centric features that encapsulate the features of all nodes belonging to the same class. This leads to including the global context of graph structure in the node feature. For each class $y_k$, we calculate a class representative $\bar{x}_k$, which is the average of the feature vectors of all nodes $v_i$ that belong to class $y_k$. This computation is formalized in Equation 4:

$$\bar{x}_k = \frac{1}{|y_k|} \sum_{v_i \in y_k} x_i \qquad (4)$$

where $|y_k|$ denotes the number of nodes in class $k$, and $x_i \in \mathbb{R}^d$ represents the feature vector of node $v_i$. The resulting class representative $\bar{x}_k$ serves as a centroid in the feature space, capturing the central tendency of the features within that class. For each node $v_i$, we calculate the cosine similarity $sim(x_i, \bar{x}_k)$ between its feature vector $x_i$ and each class representative $\bar{x}_k$. This similarity measure, defined in Equation 5, shows how closely the node's features align with the characteristics of each class.

$$sim(x_i, \bar{x}_k) = \frac{x_i \cdot \bar{x}_k}{\|x_i\| \|\bar{x}_k\|} \qquad (5)$$

After computing the similarities, the class representative that exhibits the highest similarity with the node $v_i$ is identified. This most similar class representative is then concatenated with the original feature vector of the node to produce the similarity-enriched feature matrix $X_{sim} \in \mathbb{R}^{n \times 2d}$, as described in Equation 6.

$$x_{sim}(v_i) = [X(v_i) \parallel \bar{x}_k \text{ for } k = \arg\max_k sim(x_i, \bar{x}_k)] \qquad (6)$$

This approach allows each node to inherit global information about its class, improving its feature representation by integrating not only its own attributes but also the typical characteristics of the class it is most similar to. This similarity-enriched feature matrix enhances the ability of the model to generate more correct predictions by considering both local node features and broader class-level patterns.

The final feature matrix $X_{\text{final}} \in \mathbb{R}^{n \times 3d}$ is initially formed by concatenating two distinct matrices: the connection-based convolutional feature matrix $X_{\text{deg}} \in \mathbb{R}^{n \times d}$ and the similarity-enriched feature matrix $X_{\text{sim}} \in \mathbb{R}^{n \times 2d}$. This process expands the dimensionality of the final feature matrix to encompass both local and global information, offering a robust representation that adeptly handles the complexities of graph structures. Following this expansion, $X_{\text{final}}$ is processed through a fully connected layer that compresses its dimensionality down to $d$, resulting in a streamlined final features matrix $X_{\text{final}} \in \mathbb{R}^{n \times d}$. This compression step prepares the enriched features for classification tasks. These feature vectors are subsequently utilized as inputs in the classification model, enhancing the classification performance. In the next section, we discuss the node sampling technique. The Algorithm 1 enhances node features in a graph by performing several operations sequentially. Initially, from lines 1 to 4, it aggregates weighted features from each similar node based on the Compatibility matrix, applies a non-linear transformation, and creates a convolutional feature amplification matrix $P \in \mathbb{R}^{n \times d}$. Subsequently, lines 5 to 8 compute each node's connection score and enhance its features by incorporating this score into the convolutional feature amplification matrix $P$, resulting in a connection-based convolutional feature matrix $X_{\text{deg}} \in \mathbb{R}^{n \times d}$. Lines 9 to 11 focus on deriving class-centric features for each class from the nodes within those classes. The final steps, lines 12 to 18, involve computing cosine similarities between each node's features and these class-centric features, concatenating with the most similar class-centric feature to form a similarity-driven feature matrix $X_{\text{sim}} \in \mathbb{R}^{n \times d}$, and combining $X_{\text{sim}} \in \mathbb{R}^{n \times d}$ and $X_{\text{deg}} \in \mathbb{R}^{n \times d}$ to produce the final feature matrix $X_{\text{final}} \in \mathbb{R}^{n \times d}$. This process effectively utilizes both local connectivity and global class information to refine node representations for improved performance in node classification.

---

**Algorithm 2** Sampling matrix creation

---

**Input:** Adjacency matrix $A \in \mathbb{R}^{n \times n}$, $e_i$ is a one-hot vector representing the node $i$ , c is damping factor and, $I \in \mathbb{R}^{n \times n}$ is identity matrix

**Output:** Sampling matrix $S \in \mathbb{R}^{n \times n}$

1: Initialize a matrix $D \in \mathbb{R}^{n \times n}$ with all zeros.
2: $d = A \times \mathbf{1}$
3: **for** $v_i \in \{v_1, v_2, \ldots, v_n\}$ **do**
4:     $D[v_i][v_i] = d[v_i]$
5: **end for**
6: $\hat{A} = D^{-1/2} \times A \times D^{-1/2}$
7: Compute Similarity matrix $S$
8: **for** $v_i \in \{v_1, v_2, \ldots, v_n\}$ **do**
9:     $S[v_i] = I - (1-c)\hat{A} + c \cdot e_i$
10: **end for**

---

## 4.2. Node Sampling and Sequence Construction

In traditional methods [10, 30], the Graph Transformer typically processes the entire graph, considering all nodes and edges together, which may raise scalability concerns. To address this our framework uses a node sampling technique based on the Personalized PageRank (PPR) algorithm [11] to create subgraphs around each node defined in Algorithm 3. Using the sampling matrix $S \in \mathbb{R}^{n \times n}$ where n is the number of nodes in the graph defined in Algorithm 2, we identify the most influential nodes relative to a given node $v$. The vector $r$ is defined for each node as shown in Equation 7:

$$r = c(I - (1-c)\hat{A})^{-1} e_v \tag{7}$$

where $\hat{A}$ is the symmetrically normalized adjacency matrix, calculated as,

$$\hat{A} = D^{-1/2}(A + I)D^{-1/2} \tag{8}$$

---

 

where $I \in \mathbb{R}^{n \times n}$ is the identity matrix, $D \in \mathbb{R}^{n \times n}$ is the degree matrix, $A \in \mathbb{R}^{n \times n}$ is the adjacency matrix, $e_v$ is a one-hot vector representing the node $v$, and $c$ is damping factor. The node sampling process begins by calculating scores from the sampling matrix $S \in \mathbb{R}^{n \times n}$, where each row reflects the relative influence of all nodes with respect to node $v$. To prevent the node from sampling itself, the diagonal entry is set to a large negative value. Next, the top $k_1$ neighbors of the node $v$ are identified based on their PPR scores, selecting those with the highest influence within the local neighborhood.

Following this, probabilistic sampling is performed, selecting a subset of nodes with positive PPR scores, along with the top $k_1$ neighbors, to form the subgraph. The indices of these sampled nodes, together with node $v$ itself, are then used to construct the subgraph, ensuring that each subgraph contains $k_1 + 1$ nodes. This approach effectively captures both highly connected and structurally diverse neighborhoods, allowing the model to focus on locally significant nodes while maintaining the overall graph structure. In the following module, we will explore the third component of our framework, which focuses on attention-based node classification. In Algorithm 2, lines $1 - 9$ efficiently compute the normalized adjacency matrix by Equation 8. This normalization balances node influence across the graph. This forms $\hat{A}$, a symmetric matrix that scales adjacency relations based on normalized degrees, enhancing graph structure representation. Lines $11 - 13$ then compute the sampling matrix $S$ by incorporating the normalized adjacency matrix with a damping factor $c$ and a one-hot vector for each node. It will be further used in finding the subgraph of each node in the graph explained in Algorithm 3.

---

**Algorithm 3** Subgraph sampling for ensemble

---

**Input:** Sampling matrix $S$, nodes $v_0, v_1, \ldots, v_{n-1}$, number of nodes in each subgraph $k_1$, number of samples $q$
**Output:** Subgraphs of each node

1: $data\_list \leftarrow []$
2: **for** $v_i \in \{v_0, v_1, \ldots, v_{n-1}\}$ **do**
3:      $s = S[v_i]$
4:      $s[v_i] = -\infty$
5:      $top\_neighbor\_index = s.argsort()[-k_1 :]$
6:      $s = S[v_i]$
7:      $s[v_i] = 0$
8:      $count = 0$
9:      **for** $v_j \in \{v_0, v_1, \ldots, v_{n-1}\}$ **do**
10:          $s[v_j] = \max(s[v_j], 0)$
11:          **if** $s[v_j] > 0$ **then**
12:              $count = count + 1$
13:          **end if**
14:      **end for**
15:      $r = \min(count, k_1)$
16:      $sub\_data\_list \leftarrow []$
17:      $p = s/s.sum$
18:      **for** $m \in \{1, 2, \ldots, q\}$ **do**
19:          **if** $r > 0$ **then**
20:              $node\_sample\_id = \text{selectNode}(v_i, r, p)$
21:          **else**
22:              $node\_sample\_id = []$
23:          **end if**
24:          $remaining\_nodes = \text{select}(top\_neighbor\_index, k_1 - r)$
25:          $node\_sample\_id = \text{concatenate}(v_i, node\_sample\_id, remaining\_nodes)$
26:          $sub\_data\_list.\text{append}(node\_sample\_id)$
27:      **end for**
28:      $data\_list.\text{append}(sub\_data\_list)$
29: **end for**
30: **Return** $data\_list$

---

- **selectNode**($v_i$, $sample\_index$, $p$)**:** It takes input as a node, the number of nodes to be selected, and probability vector $p$. It randomly selects $r$ number of nodes with weighted probability without repetition.

- **select**($top\_neighbor\_index$, $k_1 - r$)**:** It takes input as the $top\_neighbor\_index$ and the number of nodes to be selected. It returns randomly selected $k_1 - r$ nodes from the $top\_neighbor\_index$.

In Algorithm 3, the initialization of $data\_list$ in line 2 sets up storage for subgraphs of each node. For each node $v_i$, its entry in the sampling matrix $S$ is set to $-\infty$ in line 4 to prevent self-inclusion in subgraphs. The top $k_1$ influential neighbors are identified in line 5, and the entry for $v_i$ is reset to zero in line 6. The positive entries in $S[v_i]$ are counted between lines $8 - 12$, followed through the calculation of a probability vector $p$ for sampling in line 15. For each node, $q$ subgraphs are generated by sampling nodes using $p$ in line 18, with additional nodes selected from top neighbors to complete the subgraph size in line 22. The resulting subgraphs are stored in $data\_list$ in line 25. This approach efficiently combines matrix operations and stochastic sampling to generate localized subgraphs for each node.

### 4.3. Attention-based Node Classification
The drawbacks of Graph Transformer are discussed in this section along with solutions for better classification. Graph transformers use only the node features to calculate the attention score which ignores the structural property of the network. To address this, we introduce a structural encoding vector to incorporate structural details into the attention process.

#### 4.3.1. Structure-Aware Multi-Head Attention
Structure-Aware Multi-Head Attention (SA-MHA) is introduced to address above mentioned limitations. For a pair of nodes $v_i$ and $v_j$, multiple views of structural information are captured in a structural encoding vector, denoted as $\psi$, which enhances the attention mechanism. The attention score $\beta$ is computed as defined by Equation 9.

$$\beta = \frac{QK^T}{\sqrt{d}} + \psi \tag{9}$$

Here, $d$ represents the dimension of the key vectors, while $Q$ and $K$ denote the learnable weight matrices for the query and key, respectively. The structural encoding $\psi_{ij}$ is generated by concatenating multiple structural encoding functions represented by Equation 10.

$$\psi_{ij} = Concat(\alpha_m(v_i, v_j) \mid m \in \{0, 1, \ldots, M-1\}) \tag{10}$$

Here, $Concat$ is a concatenation operation. Each structural encoding function $\alpha_m$ captures a specific aspect of the relationship between the node pair defined in Equation 11. In our approach, we consider the structural order in which node pairs are connected. The structural encoding functions are defined in Equation 11:

$$\alpha_m(v_i, v_j) = \begin{cases} \tilde{A}_m[i,j], & \text{if } m < M-1 \\ 0, & \text{otherwise} \end{cases} \tag{11}$$

where $\tilde{A} = \text{Norm}(A + I)$ is the normalized adjacency matrix including self-connections. Thus, the first $M-1$ dimensions of $\phi_{ij}$ encode the probabilities of reachability from 0-order (identity relationship) to $(M-2)$-order between nodes $v_i$ and $v_j$, preserving fine-grained structural information for each node pair.

#### 4.3.2. Prediction
The prediction process in our framework leverages the attention mechanism to incorporate both the graph's structural information and the positional relationships between nodes. This process begins with encoding the input features $X_{\text{final}} \in \mathbb{R}^{n \times d}$ and applying a linear transformation to elevate them into a space with higher dimensionality defined in Equation 12:

$$H^{(0)} = Linear(X_{\text{final}}) \tag{12}$$

The initial encoded representation is refined using the SA-MHA, allowing the model to simultaneously capture different parts of the graph while incorporating structural biases. The query, key, and value matrices for the attention heads are computed as $Q^{(l)} = H^{(l-1)}W_Q^{(l)}$, $K^{(l)} = H^{(l-1)}W_K^{(l)}$, $V^{(l)} = H^{(l-1)}W_V^{(l)}$. The attention score for each head is computed using the scaled dot-product defined by Equation 13, enhanced by structure-aware encoding.

$$Attention(Q, K, V) = softmax \left( \frac{QK^\top}{\sqrt{d_k}} + \psi \right) V \tag{13}$$

where $\psi$ represents the attention bias, capturing both structural and positional relationships between the nodes. After the attention scores are computed, the output passes through a feed-forward network (FFN), which applies two linear transformations separated by a GELU activation function as expressed by Equation 14.

$$FFN(H^{(l)}) = GELU \left( H^{(l)}W_1 + b_1 \right) W_2 + b_2 \tag{14}$$

Residual connections and layer normalization are then applied to stabilize the training process as illustrated by Equation 15.

$$H^{(l+1)} = LayerNorm(H^{(l)} + FFN(H^{(l)})) \tag{15}$$

Finally, the hidden representation $H^{(l)}$ is used for prediction by applying a linear transformation followed by a softmax function to produce class probabilities as expressed by Equation 16.

$$T = softmax \left( H^{(L)}W_{out} + b_{out} \right) \tag{16}$$

The predicted class for each node is determined by choosing the class with the greatest probability in the output distribution $T \in \mathbb{R}^{1 \times u}$ where $u$ is a number of classes in the dataset. This method ensures that the final predictions fully use the structural and positional information embedded in the graph, allowing for accurate node classification within complex network systems.

## 5. Experiment

This section presents the datasets used for the experiments and the comparison of our proposed method with existing Graph Transformer-based and GNN-based approaches. Then we discuss the hyperparameter information and perform the ablation studies.

### 5.1. Datasets

To validate our method's adaptability we experiment on both homophilic and heterophilic datasets summarized in Table 2. Homophilic datasets include Cora, Citeseer, and Pubmed, where nodes are likely to link with similar others. Heterophilic datasets such as Wisconsin, Texas, Cornell, Chameleon, Actor, CS, Photo, and Squirrel in which nodes belonging to different classes are more prone to being connected. To understand the structural properties of the datasets we compute the following graph metrics.

- Average Degree (AD) [31]: To understand the graph structure, we calculate the average degree of nodes in each dataset. A high degree generally means a dense graph, which can then influence the learning efficiency of graph-based deep learning models.

- Clustering Coefficient (CC) [32]: This metric helps to measure the tendency of nodes to cluster together, such that get a tightly connected cluster of nodes. In general, a graph with a very high clustering coefficient has a high probability that nodes in the community will be connected.

- PageRank Centrality (PRC) [33]: We calculate the PageRank centrality to the order of incoming links to calculate the importance of each node. In particular, it is useful in node classification tasks since it can be used to identify influential nodes that should be of vital importance when information flows through the network directly affecting the classification results.

**Table 2**
Summary of homophilic and heterophilic datasets

| Dataset | # Nodes | # Edges | # Classes | # Features | Tri | AD | CC | PRC | HR |
|---------|---------|---------|-----------|------------|-----|----|----|-----|----|
| **Homophilic Datasets** | | | | | | | | | |
| Cora | 2,708 | 5,429 | 7 | 1,433 | 1.81 | 3.90 | 0.24 | 0.000369 | 0.810 |
| Citeseer | 3,327 | 4,732 | 6 | 3,703 | 1.05 | 2.74 | 0.14 | 0.000301 | 0.736 |
| PubMed | 19,717 | 44,327 | 3 | 500 | 1.90 | 4.50 | 0.060 | 0.000051 | 0.802 |
| CS | 18,333 | 163,788 | 15 | 6,805 | 14.04 | 8.93 | 0.343 | 0.000055 | 0.808 |
| Photo | 7,650 | 238,162 | 8 | 745 | 281.33 | 31.13 | 0.404 | 0.000131 | 0.827 |
| **Heterophilic Datasets** | | | | | | | | | |
| Wisconsin | 251 | 499 | 5 | 1,703 | 1.41 | 2.05 | 0.208 | 0.003984 | 0.196 |
| Texas | 183 | 295 | 5 | 1,703 | 1.10 | 1.78 | 0.198 | 0.005464 | 0.108 |
| Cornell | 183 | 280 | 5 | 1,703 | 0.97 | 1.65 | 0.167 | 0.005464 | 0.131 |
| Actor | 7,600 | 26,752 | 5 | 931 | 2.81 | 3.95 | 0.080 | 0.000132 | 0.219 |
| Squirrel | 5,201 | 198,493 | 5 | 2,089 | 5534.86 | 41.74 | 0.422 | 0.000192 | 0.224 |
| Chameleon | 2,277 | 31,421 | 5 | 2,325 | 451.99 | 15.85 | 0.481 | 0.000439 | 0.235 |

# Nodes = Number of nodes, # Edges = Number of edges, # Classes = Number of classes, # Features = Number of features per node, Tri = Triangle count, AD = Average Degree, CC = Clustering Coefficient, PRC = PageRank Centrality, HR = Homophily Ratio.

- Homophily Ratio (HR) [34]: This metric helps to measure the feature similarity of labeled nodes that are connected in a network. High symmetry ratios mean that nodes are connected with nodes in the same class, which can greatly influence the efficiency of node classification algorithms.

- Triangles (Tri) [31]: To understand how different parts of the network are connected, the probability of forming dense clusters and to better understand local connectivity, we count of triangles for each node.

### 5.1.1. Cora, Citeseer, and Pubmed
The Cora, Citeseer, and Pubmed [35] datasets are traditional citation networks that are widely adopted in graph-based tasks. In these networks, the node represents the research paper while edges connect nodes in cases where one paper cites the other. The node features are sparse bag-of-words vectors obtained from the paper textual content, and each feature indicates the presence of specific text in the paper.

### 5.1.2. CS
The Coauthor CS [36] dataset models a graph where nodes represent individual authors connected by edges if they have co-authored a paper. This forms a network reflecting real-world academic collaborations. Node features often include embeddings from their publications, such as bag-of-words from paper keywords, offering semantic insights into their research domains.

### 5.1.3. Photo
The Amazon Photo [36] dataset forms a graph where each node represents a photography-related product, linked by edges when products are frequently purchased together, mirroring consumer buying behaviors on e-commerce sites. Node features include bag-of-words extracted from product reviews, offering valuable insights into consumer preferences and the attributes of the products.

### 5.1.4. Texas and Wisconsin
The Texas and Wisconsin [37] datasets are smaller-scale web networks, frequently utilized to test model performance on challenging graph structures. In these datasets, nodes represent web pages, and edges correspond to hyperlinks between them. The node features are binary vectors indicating the presence or absence of specific keywords on the web pages.

### 5.1.5. Actor
The Actor [38] dataset, derived from the film industry, features a unique graph structure where nodes are actors and there are edges between nodes if they have collaborated professionally. The characteristics of the actors including gender, birth year, and the genres of movies they have appeared in, provide a wide range of datasets for model evaluation.

### 5.1.6. Cornell

The Cornell [39] dataset is a popular website network in which nodes are web pages and if there is a hyperlink from one page to another then there is an edge between them. Node features are the bag-of-words representation of web pages. Web pages can be classified into the five classes student, project, course, staff, and faculty.

### 5.1.7. Squirrel and Chameleon

The Squirrel and Chameleon [40] dataset are popular Wikipedia page-page network. In which nodes are the articles from Wikipedia and if there are hyperlinks between the articles then there is an edge between them. Node features indicate the presence of particular nouns in the articles. The nodes were classified into 5 classes in terms of their average monthly traffic.

## 5.2. Comparison Methods

We compare SIGNNet with the existing methods to show its effectiveness. In this section, we summarize the various methods used for comparison in our experiment. We start with GNN-based methods and conclude with the graph transformer-based methods.

### 5.2.1. GNN based Methods

We compare the performance of SIGNNet with the following GNN-based approaches.

1. GCN [3] applies a scalable approach to graphs by using a simplified version of convolutional neural networks directly on graph data. It learns node representations that capture both local structure and node features.

2. GAT [41] introduces graph attention networks, which use self-attention layers to allow nodes to focus on their most important neighbors.

3. APPNP [17] is based on GCN and PageRank to improve node classification in graphs. It is called Personalized propagation of neural predictions (PPNP), uses a propagation scheme based on Personalized PageRank and its faster version is APPNP.

4. GCNII [18] is an improved version of GCN that solves the over-smoothing problem. Two techniques are used such as initial residual connections and identity mapping. It was performing better than the existing methods present at that time.

5. ACM-GCN [42] addresses the issue of heterophily in Graph Neural Networks (GNNs), where connected nodes significantly differ, often leading to underperformance. The Adaptive Channel Mixing (ACM) framework, leverages dynamic utilization of aggregation, diversification, and identity channels in each GNN layer.

6. GATv2 [43] is an enhanced version of the original Graph Attention Network (GAT). To overcome the drawbacks of GAT it uses dynamic attention in place of static attention, which allows it to handle more complex graphs and perform better over a range of graph structures.

7. Non-local GNNs [44] introduce a new framework for graph neural networks (GNNs) that utilizes non-local aggregation, which is essential for tasks on graphs with low homophilic coefficients. The author suggests an effective attention-based technique for non-local aggregation and demonstrates how local aggregation can be harmful for certain kinds of graphs.

8. FSGNN [29] is simple and improves performance by selecting only the most useful features from graph data. L2-Normalization and SoftMax are used to eliminate less informative features that were collected from neighbors.

9. SNGNN++ [4] is a method created to overcome the difficulties in heterophilic graphs, in which nodes of the same class do not always need to be connected. To address this, a matrix is calculated to show the similarity between the nodes, helping with better aggregation and improving model performance.

10. UniG-Encoder [45] uses a projection matrix to transform node connections into edge features, which are processed by a neural network. A reverse transformation is used to derive the node embeddings, and they are used for graph-based tasks. In contrast to existing methods, UniG-Encoder seamlessly couples node features and graph structures while performing well with both heterophilic and homophilic graphs.

11. Geom-GNN [37] finds that message-passing neural networks generally do not adequately capture structural information and detect distant relationships in graphs. To deal with it, the geometric aggregation scheme employs the ideas from the network geometry to achieve worthwhile exploitation of the continuous space below the graph structure. The whole approach is based on node embedding, mapping of structural neighborhoods, and dual-level aggregation. These components are incorporated into the GeomGCN model to significantly improve direct graph learning.

### 5.2.2. Graph Transformer based Methods

We compare the performance of SIGNNet with the following Graph Transformer-based approaches.

1. SAN [10] uses learned positional encoding (LPE) to capture the full laplacian spectrum, which supports the model to understand the position of the nodes in the graph. The LPE is added to the node features, and then the fully connected transformer network processes the aggregate features. This approach offers deeper insights into the graph structure and enhances model performance.

2. UniMP [46] is a method that combines feature and label propagation for better classification from the graph convolutional network and the label propagation algorithm. It uses a graph transformer to process feature and label embeddings, and to reduce overfitting uses a masked label prediction strategy.

3. DET [26] is designed to improve the scalability in large graphs. It uses two encoders first is the structural encoder, which aggregates the information from neighbors and second is the semantic encoder, which finds semantically related nodes. This method performs better for node classification tasks.

4. Adaptive Graph Transformer AGT [31] is introduced to enhance node classification in graphs. To overcome the limitations of existing graph transformers, it incorporates trainable centrality encoding and kernelized local structure encoding to capture structural properties effectively. Additionally, it includes an adaptive transformer block, boosting performance on node classification tasks.

5. NAGphormer [47] is designed to handle large graphs efficiently. It uses a Hop2Token module that turns neighborhood features from different hops into multiple tokens for each node, instead of treating each node as a single token. It also learns more informative node representations compared to advanced Graph Neural Networks (GNNs).

6. Gapformer [25] is designed to improve node classification using Graph Transformers. It focuses on two issues: the first is irrelevant information from distant nodes, and the second is high computational cost. This method overcomes these issues by using graph pooling, which reduces the number of nodes for attention while keeping long-range information, thus reducing complexity.

7. SoftGNN [48] Graph neural networks excel in networks where similar nodes connect but struggle in heterophilic networks with dissimilar connections. Introduces a label-guided GNN that leverages node labels to selectively aggregate neighborhood information across different classes. By incorporating an adaptive attention mechanism, this model enhances node representation distinguishability and generalizes better across varying network types.

## 5.3. Experimental Results

In this section, we present a comprehensive overview of the experimental setup, describing the datasets, model configurations, hyperparameter settings, and evaluation metrics. Following this, we discuss the performance comparison on both homophilic and heterophilic datasets.

**Table 3**
Comparison of models on homophily datasets

| Model | Cora | Citeseer | Pubmed | CS | Photo |
|---|---|---|---|---|---|
| **Graph Neural Network-based Methods** | | | | | |
| GCN [3] | 86.92±1.33 | 76.13±1.51 | 88.42±0.50 | 89.11±0.70 | 85.94±1.18 |
| GAT [41] | 87.34±1.14 | 75.75±1.86 | 86.33±0.48 | 88.53±0.54 | 87.13±1.00 |
| APPNP [17] | 87.75±1.30 | 76.53±1.61 | 86.52±0.61 | 94.49±0.07 | 94.32±0.14 |
| GCNII [18] | 86.08±2.18 | 74.75±1.76 | **90.15±0.43** | 84.23±0.78 | 67.06±1.74 |
| ACM-GCN [42] | 87.91±0.95 | 77.41±1.78 | 90.30±0.52 | 94.83±0.24 | 90.34±1.82 |
| GATv2 [43] | 87.25±0.89 | 75.72±1.30 | 85.75±0.55 | 88.46±0.61 | 81.52±3.23 |
| Non-local GNNs [44] | 88.50±1.80 | 76.20±1.60 | 89.06±0.27 | 94.28±1.82 | 94.18±0.19 |
| FSGNN [29] | 88.23±1.17 | 77.40±1.93 | 89.78±0.38 | 95.15±0.48 | 92.58±0.78 |
| SNGNN++ [4] | 88.13±0.98 | 77.75±1.75 | 87.23±0.42 | 92.26±1.28 | 93.67±0.97 |
| UniG-Encoder [45] | 88.49±1.10 | 77.78±1.70 | 89.76±0.46 | 94.57±1.93 | 92.92±1.27 |
| Geom-GNN [37] | 85.35±1.57 | 78.90±1.15 | 89.95±0.47 | 95.64±2.81 | 93.35±1.40 |
| **Graph Transformer-based Methods** | | | | | |
| SAN [10] | 81.91±3.42 | 69.63±3.76 | 81.79±0.98 | 94.51±0.15 | 94.86±0.10 |
| UniMP [46] | 84.18±1.39 | 75.00±1.59 | 88.56±0.32 | 94.20±0.34 | 92.49±0.47 |
| DET [31] | 86.30±1.41 | 75.37±1.41 | 86.28±0.44 | 93.34±0.31 | 91.44±0.49 |
| AGT [31] | 81.65±0.41 | 70.95±0.62 | 81.48±1.69 | 87.16±2.18 | 88.49±0.92 |
| NAGphormer [47] | 85.77±1.35 | 73.69±1.48 | 89.70±0.19 | 95.75±0.09 | **95.49±0.11** |
| Gapformer [25] | 87.37±0.76 | 76.21±1.47 | 88.49±0.44 | 94.48±0.36 | 92.34±0.63 |
| SoftGNN [48] | 86.34±0.85 | 76.91±1.57 | 89.45±0.35 | 95.38±1.27 | 92.17±0.54 |
| **SIGNNet** | **92.45±1.13** | **81.14±0.52** | 87.89±2.10 | **97.36±0.81** | 94.84±0.61 |

### 5.3.1. Experimental Setup

To evaluate the performance of the SIGNNet, we run this model using the Adam optimizer for the GCN with a learning rate of 0.01 and a weight decay of 5e-4. The hidden dimensions are set to 128, and dropout of 0.3 to avoid overfitting. We used the AdamW optimizer for the Graph Transformer model with an initial learning rate of 2e-4 and an ending learning rate of 1e-9. The hidden dimensions are set to 128 with a dropout of 0.3. The weight decay is 0.01, and the batch size of 32. To evaluate the performance of SIGNNet, we use accuracy as an evaluation metric. We implement our method using PyTorch Geometric (PyG) [49] and PyTorch [50]. The datasets are publicly available on PyG, and we use the same node and label attributes as provided in these datasets. We divide each dataset into 60% for training, 20% for validation, and 20% for testing to maintain consistent evaluation across the model.

### 5.3.2. Performance on Homophilic Datasets

From the results presented in Table 3, it is clear that SIGNNet achieves state-of-the-art performance on three out of the five datasets, which underscores the effectiveness of our proposed method. Our analysis compares methods based on both graph neural networks and graph transformers. Unlike Non-local GNNs [44], which mainly focus on heterophilic datasets and tend to overlook local network information, our method enhances node features by aggregating both structural and semantic information of the network. This is achieved through the use of connection scores and class-centric features, leading to improved performance. Specifically, the accuracy of Non-local GNNs on the Cora, Citeseer, Pubmed, CS, and Photo datasets are 88.50%, 76.20%, 89.06%, 94.28%, and 94.18% respectively. In contrast,

SIGNNet surpasses Non-local GNNs on four out of these five datasets, with improvements of 3.95%, 4.94%, 3.08%, and 0.66% for Cora, Citeseer, CS, and Photo respectively. These results demonstrate the enhanced capability of SIGNNet in handling datasets where similar nodes are more likely to be connected, thereby providing a more accurate classification performance. AGT [31] demonstrates strong performance in node classification but encounters scalability challenges because it processes the entire graph for each classification task. To address this, our proposed method employs a sampling technique that not only mitigates scalability issues but also emphasizes the importance of neighboring nodes. This focus enhances the understanding of local node interactions and improves performance on homophilic datasets, where similar nodes are more likely to be connected. The accuracies of AGT on the Cora, Citeseer, Pubmed, CS, and Photo datasets are 81.65%, 70.95%, 81.48%, 87.16%, and 88.49%, respectively. In contrast, SIGNNet outperforms AGT across all five datasets with significant improvements. Specifically, SIGNNet achieves enhancements of 10.80% for Cora, 10.19% for Citeseer, 6.41% for Pubmed, 10.20% for CS, and 6.35% for Photo. These results highlight SIGNNet's superior capability in handling diverse dataset characteristics effectively.

### 5.3.3. Performance on Heterophilic Datasets

From the results presented in Table 4, it is evident that SIGNNet achieves state-of-the-art performance on five of the six datasets, underscoring the robustness of our proposed method. FSGNN [29] primarily utilizes local features and neglects the structural property of the network. This approach generally results in diminished performance on heterophilic graphs, where nodes belonging to different classes are more likely to be connected. To address this limitation, our method incorporates a compatibility matrix that facilitates information sharing among similar nodes and integrates class-centric features to provide a comprehensive view of each class. This dual strategy enhances our understanding of the network and significantly improves performance. The accuracies of FSGNN on the Wisconsin, Texas, Actor, Cornell, Chameleon, and Squirrel datasets are 88.43%, 87.57%, 35.75%, 87.84%, 78.95%, and 74.10%, respectively. In stark contrast, SIGNNet demonstrates superior performance over FSGNN on five of these six datasets, with accuracy improvements of 1.77% for Wisconsin, 4.67% for Texas, 4.07% for Actor, 3.51% for Cornell, and 5.02% for Chameleon. These enhancements highlight SIGNNet's effective adaptation to the challenges posed by heterophilic datasets and its ability to leverage both local and global network properties for improved classification results. Gapformer [25] faces challenges like quadratic complexity and noise from irrelevant nodes in node classification, which SIGNNet addresses effectively. SIGNNet reduces complexity through a Personalized PageRank-based node sampling method that limits the number of nodes processed. It also incorporates a Structure-Aware Multi-Head Attention (SA-MHA) mechanism, which integrates structural information into the attention process, focusing on topologically significant nodes and reducing noise from irrelevant ones. Additionally, SIGNNet enhances node representation by employing connection scores and class-centric features, allowing for a better capture of both local and global graph properties. These improvements enable SIGNNet to handle the graph structure more adeptly, thereby improving its performance in node classification tasks across various datasets. The accuracies of Gapformer on the Wisconsin, Texas, Actor, Cornell, Chameleon, and Squirrel datasets are 83.53%, 80.27%, 36.90%, 77.57%, 81.04%, and 70.98%, respectively. In clear contrast, SIGNNet significantly outperforms Gapformer across all these datasets, demonstrating notable improvements. Specifically, SIGNNet achieves enhancements of 6.67% for Wisconsin, 11.97% for Texas, 2.92% for Actor, 13.78% for Cornell, 2.93% for Chameleon, and 2.51% for Squirrel, underlining its superior performance across a range of graph environments. Furthermore, SIGNNet is compared against NAGphormer[47], SoftGNN [48], SAN [10], and many other advanced methods in node classification. SIGNNet consistently outperforms these methods by significant margins across all evaluation metrics. This thorough comparison emphasizes the robustness and effectiveness of SIGNNet specifically in node classification tasks. By surpassing existing benchmarks, SIGNNet proves its capability to effectively utilize graph structural and feature information, thus improving node classification performance across diverse datasets.

**Table 4**
Comparison of models on heterophily datasets

| Model | Wisconsin | Texas | Actor | Cornell | Chameleon | Squirrel |
|---|---|---|---|---|---|---|
| Graph Neural Network-based Methods | | | | | | |
| GCN [3] | 52.55±4.27 | 60.81±8.03 | 28.73±1.17 | 60.54±5.30 | 64.82±2.24 | 53.43±2.01 |
| GAT [41] | 57.45±3.51 | 62.16±4.52 | 28.33±1.13 | 61.89±5.05 | 60.26±2.50 | 40.72±1.55 |
| APPNP [17] | 55.29±3.90 | 61.62±5.37 | 29.42±0.81 | 73.51±0.37 | 54.30±0.77 | 32.37±0.51 |
| GCNII [18] | 52.54±7.32 | 58.91±4.32 | 25.40±0.97 | 77.86±3.79 | 63.56±3.04 | 38.47±1.58 |
| ACM-GCN [42] | 88.28±3.64 | 87.99±4.64 | 36.08±1.03 | 85.14±6.07 | 66.93±1.85 | 54.40±1.88 |
| GATv2 [43] | 52.74±3.96 | 60.54±4.55 | 28.79±1.47 | 50.27±8.97 | 62.20±2.11 | 50.80±3.01 |
| Non-local GNNs [44] | 87.30±4.30 | 85.40±3.80 | 37.90±1.30 | 84.90±5.70 | 70.10±2.90 | 59.00±1.20 |
| FSGNN [29] | 88.43±3.22 | 87.57±4.86 | 35.75±0.96 | 87.84±6.19 | 78.95±0.86 | 74.10±1.89 |
| SNGNN++ [4] | 89.02±3.36 | 88.65±4.90 | 34.20±1.26 | 87.84±4.81 | 80.83±0.94 | **75.48±1.98** |
| UniG-Encoder [45] | 87.84±3.90 | 85.95±3.90 | 35.79±0.39 | 86.75±6.56 | 81.06±1.21 | 74.39±2.17 |
| Geom-GNN [37] | 64.51±3.58 | 66.76±2.78 | 31.46±1.25 | 60.54±3.67 | 60.00±2.81 | 38.15±0.92 |
| Graph Transformer-based Methods | | | | | | |
| SAN [10] | 51.37±3.08 | 60.17±6.66 | 27.12±2.59 | 50.85±8.54 | 48.74±5.25 | 46.36±3.27 |
| UniMP [46] | 79.60±5.41 | 73.51±8.44 | 35.15±0.84 | 66.48±12.5 | 69.17±0.18 | 65.84±2.37 |
| DET [26] | 54.90±6.56 | 56.76±4.98 | 28.94±0.64 | 72.18±7.14 | 67.94±3.71 | 64.26±0.87 |
| AGT [31] | 83.20±4.10 | 79.14±4.29 | 35.74±0.51 | 73.28±4.82 | 75.83±3.84 | 73.67±3.84 |
| NAGphormer [47] | 62.55±6.22 | 63.51±6.53 | 34.33±0.94 | 56.22±8.08 | 75.17±3.85 | 68.27±3.91 |
| Gapformer [25] | 83.53±3.42 | 80.27±4.01 | 36.90±0.82 | 77.57±3.43 | 81.04±1.65 | 70.98±2.31 |
| SoftGNN [48] | 88.63±3.37 | 88.11±4.39 | 36.68±0.86 | 78.92±3.78 | 81.67±1.81 | 74.16±0.39 |
| **SIGNNet** | **90.20±1.66** | **92.24±2.81** | **39.82±1.25** | **91.35±0.48** | **83.97±0.37** | 73.49±0.51 |

## 5.4. Ablation Studies

To show the effectiveness of individual components employed in our proposed method, we perform ablation studies by removing one component at a time from our proposed framework and evaluate how each contributes to performing on three datasets: Cora, Citeseer, and Actor. It provides insight into the role any individual component plays in the feature representation and the classification performance of the model. Additionally, we assess the behavior of the model with a different number of epochs, visualize learned embeddings, and examine the effects of varying transformer layers and subgraph counts to gain insights into the adaptability and effectiveness of the framework.

### 5.4.1. Modal Component Analysis

To understand the importance of each module in the framework, we remove one module at a time and investigate its effect on the framework, comparing the results on the homophily and heterophily datasets as shown in Table 5 and Table 6. Next, we discuss the effect of each module in the framework.

**a) Impact of Neighborhood Influence Learning Feature:** It works by using message passing to add local features from neighboring nodes into each node's feature, making the node features more informative. By including this local information, we enhance the overall understanding of the node's position in the network. If we remove this module, the node features will not include any local context.

**b) Impact of Connection-aware Feature Representation:** We incorporate broader information into each node's feature by adding its connection score, which measures how influential a node is within the network. Connection score helps capture the importance of a node in its immediate surroundings. If this part is removed, the node features will lose their global context, reducing their ability to reflect the node's influence within the network.

**c) Impact of Class-centric Feature Representation:** We incorporate global information into each node by aggregating class-centric features. The class-centric information is calculated by taking the average of the feature vectors for all nodes that belong to the same class. If we remove this module, the node features will be missing the broader contextual understanding of the graph structure, and this would negatively impact the method's performance.

**Table 5**
Performance on homophilic datasets across different configurations

| Configuration | Cora | Citeseer | Pubmed | CS | Photo |
|---|---|---|---|---|---|
| W/o GCN | 87.45 | 74.44 | 83.54 | 94.64 | 93.74 |
| W/o $C_{score}$ | 91.95 | 80.97 | 87.35 | 96.81 | 94.17 |
| W/o CR | 90.77 | 78.79 | 86.14 | 96.24 | 93.88 |
| **SIGNNet** | **92.45** | **81.14** | **87.89** | **97.36** | **94.84** |

W/o GCN = Without graph convolutional network, W/o $C_{score}$ = Without connection score, W/o CR = Without class representative.

**Table 6**
Performance on heterophilic datasets across different configurations

| Configuration | Actor | Wisconsin | Texas | Cornell | Chameleon | Squirrel |
|---|---|---|---|---|---|---|
| W/o GCN | 36.15 | 84.79 | 87.05 | 86.84 | 80.28 | 70.68 |
| W/o $C_{score}$ | 38.14 | 89.57 | 90.87 | 91.15 | 83.17 | 72.78 |
| W/o CR | 37.15 | 89.46 | 91.89 | 89.75 | 83.21 | 73.29 |
| **SIGNNet** | **39.82** | **90.20** | **92.24** | **91.35** | **83.97** | **73.49** |

W/o GCN = Without graph convolutional network, W/o $C_{score}$ = Without connection score, W/o CR = Without class representative.

### 5.4.2. Performance Comparison across Various Layers

In this subsection, we evaluate the classification performance of our method by varying the number of transformer layers from 1 to 5. As the number of layers increases, the transformer captures more structural information from the network, generally improving classification performance. However, it can also cause overfitting. For example, Cora and Citeseer achieve better results with 3 layers, while the Actor dataset performs better with 2 layers. Thus, the selection of layers must be done carefully, taking into account the structural characteristics of the dataset for better performance. The following graphs Figure 4a, Figure 4b, and Figure 4c shows how the number of layers affects the performance of the Cora, Citeseer, and Actor datasets respectively.

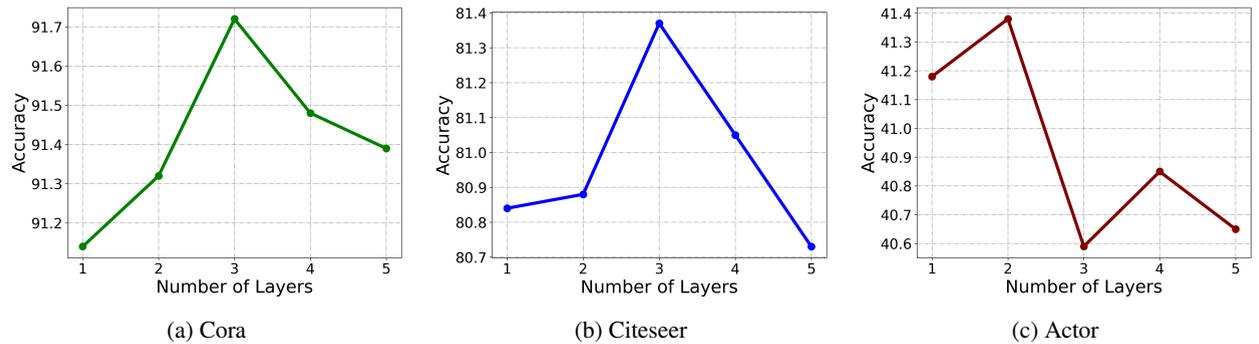

(a) Cora  (b) Citeseer  (c) Actor

**Figure 4:** Performance analysis across varying transformer layers on Cora, Citeseer, and Actor datasets.

### 5.4.3. Performance Comparison across Various Subgraphs

In this section, we analyze how changing the number of subgraphs affects the SIGNNet performance. The model exhibits the lowest performance with a single subgraph, while performance steadily improves as the number of subgraphs increases, reaching its peak at five subgraphs. Further increases lead to a decline in performance. Figure 5a, Figure 5b, and Figure 5c display the results for subgraph counts of 1, 3, 4, and 7 across the Cora, Citeseer, and Actor datasets respectively, highlighting significant variations in performance.

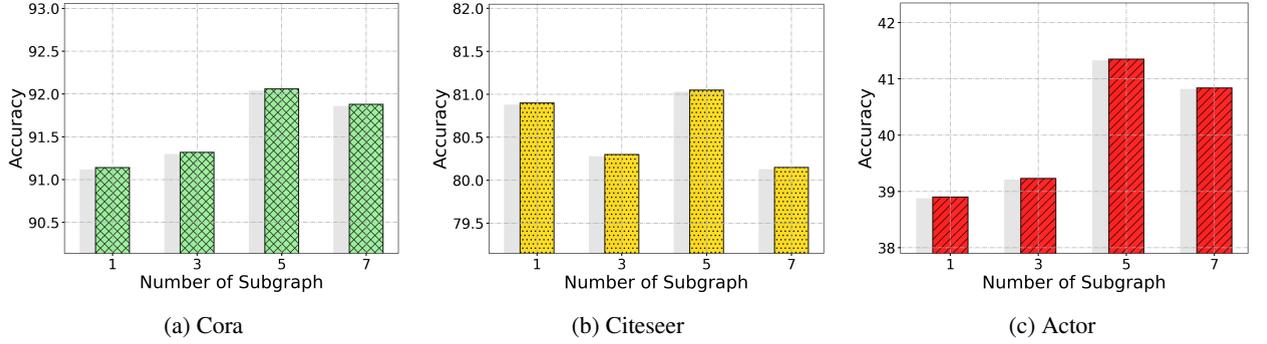

**Figure 5:** Performance analysis with varying subgraph counts for Cora, Citeseer, and Actor datasets.

### 5.4.4. Performance Comparison across Various values of Damping Factor

We analyze the role of the damping factor in the Personalized PageRank (PPR) algorithm and its impact on model performance across different datasets. The restart probability controls the balance between local and global exploration in a graph. To guarantee a bias toward nodes nearby, the random walker either returns to a predetermined starting node at each step or keeps investigating nearby nodes. When the value of restart probability is small then it encourages a more comprehensive examination of the graph, and when the number is high then it more focuses on local neighbors. Cora's performance improves with a higher restart chance as shown in Figure 6a, and Citeseer gives the best performance when the value is 0.15 as shown in Figure 6b. The Actor's accuracy, on the other hand, varies, suggesting a more intricate link as shown in Figure 6c. These results show that the damping factor is an important parameter.

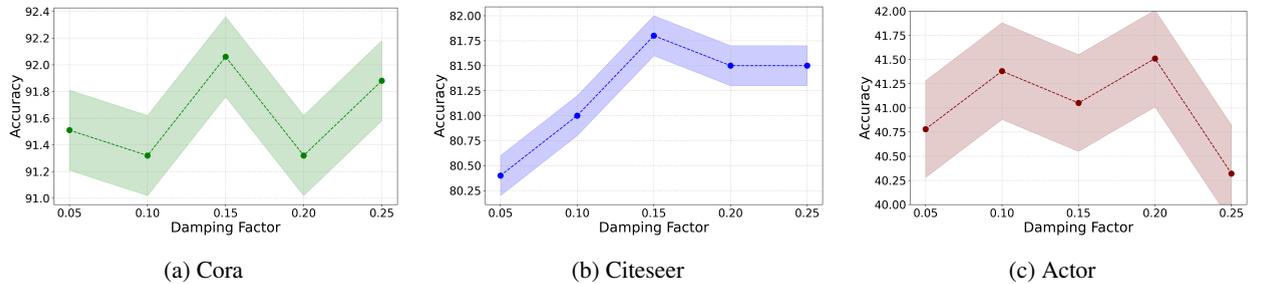

**Figure 6:** Performance analysis with varying damping factor value for Cora, Citeseer, and Actor datasets.

### 5.4.5. Performance Comparison across Various Numbers of Attention Heads

To evaluate the performance gain achieved by employing different number of attention heads. We carried our experiments on three benchmark datasets such as Cora, Citeseer, and Actor to show the significance. The number of attention heads in a graph transformer is crucial because it enables the model to focus on various node-to-node interaction characteristics by capturing several attention patterns at once. The large number of attention heads helps to capture complex patterns and identify more intricate dependencies within the graph. However, too many heads can raise computational overhead and model complexity, which could result in overfitting or performance degradation. The accuracy varies with the number of attention heads, highlighting dataset-specific trends. We observe a different trend in all the datasets as the number of attention heads increases. In Cora, a general trend is observed, the performance of the model increases as the number of heads increases as shown in Figure 7a, while Citeseer reaches its peak performance at a specific value before stabilizing as shown in Figure 7b. Due to the dense nature of Actor dataset has a more complex pattern, where accuracy increases at a smaller number of heads and then gradually declines with larger values as shown in Figure 7c. These results indicate that the optimal number of attention heads is dataset-dependent and is required to achieve good performance.

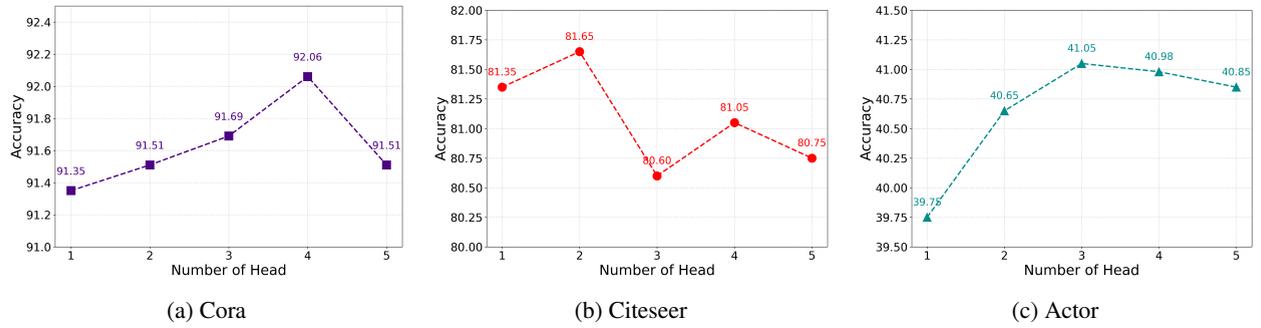

**Figure 7:** Performance analysis with varying number of heads in graph transformer for Cora, Citeseer, and Actor datasets.

### 5.4.6. Accuracy Comparison Over Epochs

In this subsection, we illustrate the performance of the method over time. In the initial phase of learning on the Cora dataset, the training accuracy increases rapidly as the model understands the dataset. Validation accuracy is also increasing but at a slower pace shown in Figure 8a. As the training proceeds, both the training and validation accuracies converge, indicating that the model learns well and does not overfit. This trend indicates that the model handles the dataset well and achieves good classification results after several epochs. From Figure 8b, we can clearly see that the Citeseer dataset is a little bit more complex than the Cora dataset as during the training validation accuracy is less steady, going up and down across the epochs, which shows that the model has a harder time generalizing to new data compared to Cora dataset. The Actor dataset is a heterophilic graph, and the Figure 8c clearly shows how difficult it is to find nodes correctly for the model. Both training and validation accuracy start at low points, while training accuracy improves steadily, validation accuracy changes do not fully stabilize, reflecting the difficulties of the Actor dataset, where the model is challenged to always achieve better results.

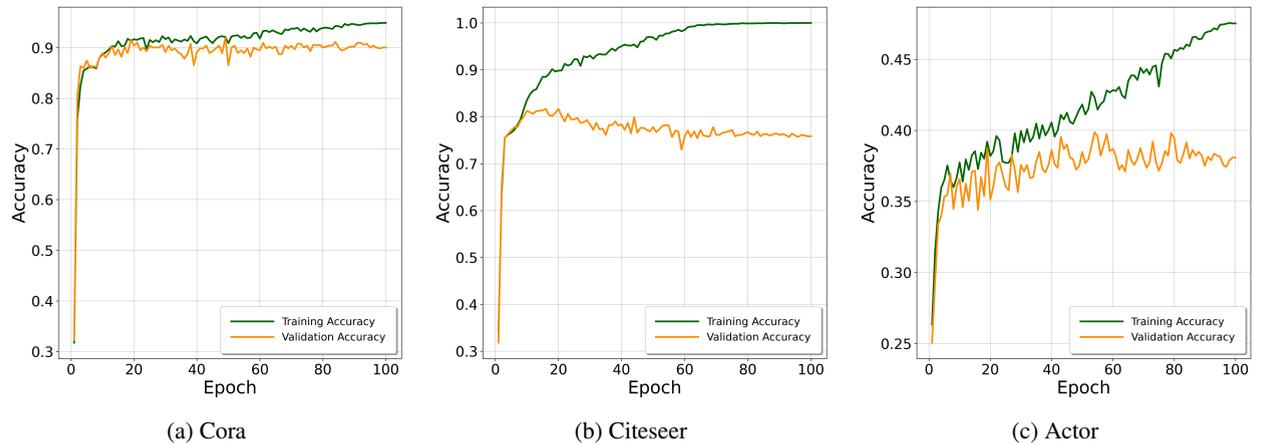

**Figure 8:** Test and validation accuracy over epochs for Cora, Citeseer, and Actor dataset.

### 5.4.7. Cora Dataset Visualization

In this subsection, we present a wide view of the Cora dataset after the model has learned through structural and feature-based transformations. This visualization captures the initial representation of the node embeddings before training shown in Figure 9a, showing minimal separation between classes. After training with the proposed method, the node embeddings are improved by capturing both the structure and semantic contexts of the network, highlighting how the method enhances the embeddings such that nodes that belong to the same class are closer to each other shown in Figure 9b. This progression illustrates the method's ability to effectively organize and classify the nodes, leading to more accurate node classification.

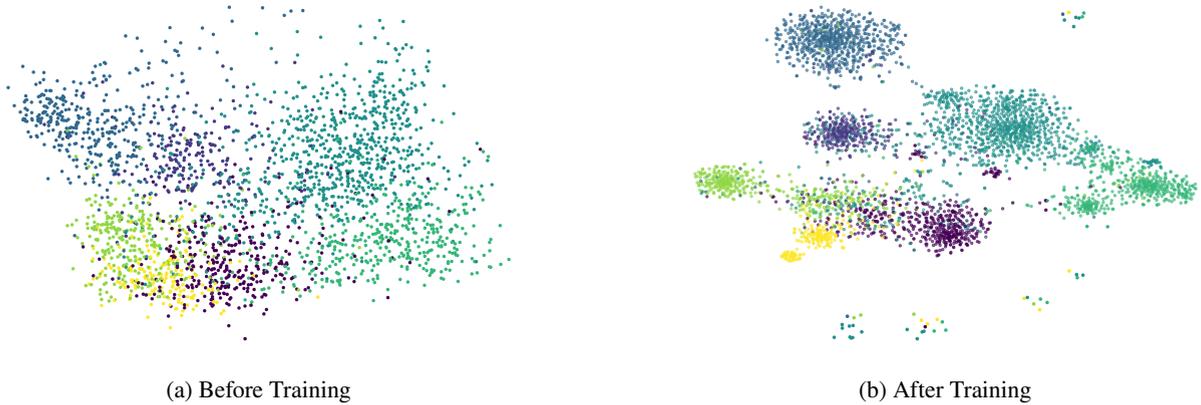

(a) Before Training            (b) After Training

**Figure 9:** Cora dataset visualization before and after training.

## 6. Discussion

We present SIGNNet, a novel framework designed for node classification by integrating both semantic and structural properties of the network into the node features. SIGNNet leverages a Personalized PageRank algorithm to generate subgraphs specific to each node which addresses scalability issues associated with Graph Transformers. Additionally, we propose a structure-aware mechanism that aggregates the structural information in the attention process of the network. Our proposed method offers significant advancements in node classification by employing some novel techniques.

### 6.1. Limitations and Future Work

The proposed framework shows promising outcomes. However, some technical limitations exist and are described below.

- Dependency on Labeled Graphs: This method is not suitable for unsupervised and semi-supervised tasks as we are dealing with the dataset in which nodes are labeled.

- Partial Loss of Global Structural Information: The node sampling method preserves key features for efficiency while improving the scalability of the graph transformer with little effect on larger network information.

- Application to Other Graph Tasks: SIGNNet focuses only on node classification tasks. This technique, however, can be expanded to additional graph tasks like community detection, link prediction, and graph classification.

- Extension to Hypergraphs: As this method is currently applied only to homogeneous and heterogeneous graphs, we can extend our analysis to more complex graphs such as hypergraphs.

Further research and experiments can overcome these limitations and make the proposed method effective and suitable for real-world node classification tasks.

## 7. Conclusion

We present a novel framework SIGNNet for node classification which integrates the structural and semantic information of the network into the node features containing only the attributes details. The node features get enriched through neighborhood-influenced feature learning, connection-aware, and class-centric feature representation. Neighborhood-influenced feature learning leverages GCNs to incorporate local structural information into node representations, enabling the model to better distinguish between classes based on local patterns. Class-centric and connection-aware features capture broader information of the network and node importance, resulting in more comprehensive and discriminative feature representations. To manage the scalability of Graph Transformers, we implement a Personalized PageRank-based node sampling method that selects the most influential nodes in the subgraph, reducing computational load while maintaining performance. Additionally, we introduce a Structure-Aware Multi-Head Attention mechanism, which incorporates both feature similarity and graph structure information, resulting in improved performance in the attention mechanism. Our framework exhibits superior performance against state-of-the-art methods. SIGNNet demonstrates better performance compared to Gapformer across both homophilic and heterophilic datasets. In the context of homophily, SIGNNet outperforms Gapformer by 5.08% on Cora, 4.93% on Citeseer, 2.88% on CS, and 2.5% on Photo. Meanwhile, within heterophilic datasets, SIGNNet exceeds Gapformer's performance by 6.67% on Wisconsin, 11.97% on Texas, 2.92% on Actor, 13.78% on Cornell, 2.93% on Chameleon, and 2.51% on Squirrel. It also exceeds SNGNN++ by 4.32% on Cora, 3.39% on Citeseer, 0.66% on Pubmed 5.10% on CS, and 1.17% on Photo. Meanwhile, within heterophilic datasets, SIGNNet exceeds SNGNN++ performance by 1.18% on Wisconsin, 3.59% on Texas, 5.62% on Actor, 3.51% on Cornell, and 3.14% on Chameleon. These results highlight the effectiveness of our approach. SIGNNet enhances node classification by combining local and global contextual information within node features. This approach boosts classification performance and tackles scalability challenges in graph transformers, providing an efficient and robust solution for node classification tasks.

**Acknowledgement**

We are thankful to the Young Faculty Research Catalysing Grant (YFRCG) scheme, an initiative by IIT Indore, for providing a research grant to Dr. Nagendra Kumar and Dr. Ranveer Singh (Project ID: IITI/YFRCG/2023-24/03).

# References

[1] J. Zhou, G. Cui, S. Hu, Z. Zhang, C. Yang, Z. Liu, L. Wang, C. Li, M. Sun, Graph neural networks: A review of methods and applications (2020). doi:10.1016/j.aiopen.2021.01.001.

[2] G. Nikolentzos, M. Vazirgiannis, Learning structural node representations using graph kernels, IEEE Transactions on Knowledge and Data Engineering 33 (2021). doi:10.1109/TKDE.2019.2947478.

[3] T. N. Kipf, M. Welling, Semi-supervised classification with graph convolutional networks, in: 5th International Conference on Learning Representations, ICLR 2017 - Conference Track Proceedings, 2017.

[4] M. Zou, Z. Gan, R. Cao, C. Guan, S. Leng, Similarity-navigated graph neural networks for node classification, Information Sciences 633 (2023). doi:10.1016/j.ins.2023.03.057.

[5] S. Yuan, H. Zeng, Z. Zuo, C. Wang, Overlapping community detection on complex networks with graph convolutional networks, Computer Communications 199 (2023) 62–71. doi:https://doi.org/10.1016/j.comcom.2022.12.008.

[6] B. Li, D. Pi, Learning deep neural networks for node classification, Expert Systems with Applications 137 (2019) 324–334. doi:https://doi.org/10.1016/j.eswa.2019.07.006.

[7] Y. Xie, Y. Liang, M. Gong, A. K. Qin, Y. S. Ong, T. He, Semisupervised graph neural networks for graph classification, IEEE Transactions on Cybernetics 53 (2023). doi:10.1109/TCYB.2022.3164696.

[8] J. H. Giraldo, K. Skianis, T. Bouwmans, F. D. Malliaros, On the trade-off between over-smoothing and over-squashing in deep graph neural networks, in: International Conference on Information and Knowledge Management, Proceedings, 2023, pp. 566–576. doi:10.1145/3583780.3614997.

[9] S. Yun, M. Jeong, S. Yoo, S. Lee, S. S. Yi, R. Kim, J. Kang, H. J. Kim, Graph transformer networks: Learning meta-path graphs to improve gnns, Neural Networks 153 (2022). doi:10.1016/j.neunet.2022.05.026.

[10] D. Kreuzer, D. Beaini, W. Hamilton, V. Létourneau, P. Tossou, Rethinking graph transformers with spectral attention, in: M. Ranzato, A. Beygelzimer, Y. Dauphin, P. Liang, J. W. Vaughan (Eds.), Advances in Neural Information Processing Systems, Vol. 34, Curran Associates, Inc., 2021, pp. 21618–21629. doi:10.48550/arXiv.2410.11189.

[11] M. Yang, H. Wang, Z. Wei, S. Wang, J. R. Wen, Efficient algorithms for personalized pagerank computation: A survey, IEEE Transactions on Knowledge and Data Engineering 36 (2024). doi:10.1109/TKDE.2024.3376000.

[12] T. Zhao, X. Zhang, S. Wang, Graphsmote: Imbalanced node classification on graphs with graph neural networks, in: Proceedings of the 14th ACM International Conference on Web Search and Data Mining, WSDM '21, Association for Computing Machinery, New York, NY, USA, 2021, p. 833–841. doi:10.1145/3437963.3441720.

[13] B. Molokwu, S. B. Shuvo, N. C. Kar, Z. Kobti, Node classification and link prediction in social graphs using rlvecn, in: Proceedings of the 32nd International Conference on Scientific and Statistical Database Management, SSDBM '20, Association for Computing Machinery, New York, NY, USA, 2020, p. 833–841. doi:10.1145/3400903.3400928.

[14] G. Li, M. Muller, G. Qian, I. C. Delgadillo, A. Abualshour, A. Thabet, B. Ghanem, Deepgcns: Making gcns go as deep as cnns, IEEE Transactions on Pattern Analysis and Machine Intelligence 45 (2023). doi:10.1109/TPAMI.2021.3074057.

[15] R. Burkholz, J. Quackenbush, D. Bojar, Using graph convolutional neural networks to learn a representation for glycans, Cell Reports 35 (2021). doi:10.1016/j.celrep.2021.109251.

[16] S. Xiao, S. Wang, Y. Dai, W. Guo, Graph neural networks in node classification: survey and evaluation, Machine Vision and Applications 33 (2022). doi:10.1007/s00138-021-01251-0.

[17] J. Klicpera, A. Bojchevski, S. Gunnemann, Combining neural networks with personalized pagerank for classification on graphs, arXiv (2018). doi:10.48550/arXiv.1810.05997.

[18] M. Chen, Z. Wei, Z. Huang, B. Ding, Y. Li, Simple and deep graph convolutional networks, in: H. D. III, A. Singh (Eds.), Proceedings of the 37th International Conference on Machine Learning, Vol. 119 of Proceedings of Machine Learning Research, PMLR, 2020, pp. 1725–1735. doi:https://doi.org/10.48550/arXiv.2007.02133.

[19] A. J. Fofanah, D. Chen, L. Wen, S. Zhang, Addressing imbalance in graph datasets: Introducing gate-gnn with graph ensemble weight attention and transfer learning for enhanced node classification, Expert Systems with Applications 255 (2024) 124602. doi:https://doi.org/10.1016/j.eswa.2024.124602.

[20] H. Shomer, Y. Ma, H. Mao, J. Li, B. Wu, J. Tang, Lpformer: An adaptive graph transformer for link prediction, in: Proceedings of the 30th ACM SIGKDD Conference on Knowledge Discovery and Data Mining, KDD '24, Association for Computing Machinery, New York, NY, USA, 2024, p. 2686–2698. doi:10.1145/3637528.3672025.

[21] M. Li, Y. Zhang, W. Zhang, S. Zhao, X. Piao, B. Yin, Csat: Contrastive sampling-aggregating transformer for community detection in attribute-missing networks, IEEE Transactions on Computational Social Systems 11 (2024). doi:10.1109/TCSS.2023.3292145.

[22] G. N. Chandrika, K. Alnowibet, K. S. Kautish, E. S. Reddy, A. F. Alrasheedi, A. W. Mohamed, Graph transformer for communities detection in social networks, Computers, Materials and Continua 70 (2022). doi:10.32604/cmc.2022.021186.

[23] S. S. Dar, M. Z. U. Rehman, K. Bais, M. A. Haseeb, N. Kumar, A social context-aware graph-based multimodal attentive learning framework for disaster content classification during emergencies, Expert Systems with Applications 259 (2025) 125337. doi:https://doi.org/10.1016/j.eswa.2024.125337.

[24] S. S. Dar, M. K. Karandikar, M. Z. U. Rehman, S. Bansal, N. Kumar, A contrastive topic-aware attentive framework with label encodings for post-disaster resource classification, Knowledge-Based Systems 304 (2024) 112526. doi:https://doi.org/10.1016/j.knosys.2024.112526.

[25] C. Liu, Y. Zhan, X. Ma, L. Ding, D. Tao, J. Wu, W. Hu, Gapformer: Graph transformer with graph pooling for node classification, in: E. Elkind (Ed.), Proceedings of the Thirty-Second International Joint Conference on Artificial Intelligence, IJCAI-23, International Joint Conferences on Artificial Intelligence Organization, 2023, pp. 2196–2205, main Track. doi:10.24963/ijcai.2023/244.

[26] L. Guo, Q. Zhang, H. Chen, Unleashing the power of transformer for graphs, arXiv preprint arXiv:2202.10581 (2022). doi:10.48550/arXiv.2202.10581.

[27] Z. Huang, Y. Tang, Y. Chen, A graph neural network-based node classification model on class-imbalanced graph data, Knowledge-Based Systems 244 (2022). doi:10.1016/j.knosys.2022.108538.

[28] J. Wu, J. He, J. Xu, Demo-net: Degree-specific graph neural networks for node and graph classification, in: Proceedings of the ACM SIGKDD International Conference on Knowledge Discovery and Data Mining, 2019, pp. 406–415. doi:10.1145/3292500.3330950.

[29] S. K. Maurya, X. Liu, T. Murata, Simplifying approach to node classification in graph neural networks, Journal of Computational Science 62 (2022). doi:10.1016/j.jocs.2022.101695.

[30] C. Ying, T. Cai, S. Luo, S. Zheng, G. Ke, D. He, Y. Shen, T. Y. Liu, Do transformers really perform bad for graph representation?, in: Advances in Neural Information Processing Systems, Vol. 34, 2021, pp. 28877–28888.

[31] X. Ma, Q. Chen, Y. Wu, G. Song, L. Wang, B. Zheng, Rethinking structural encodings: Adaptive graph transformer for node classification task, in: Proceedings of the ACM Web Conference 2023, WWW '23, Association for Computing Machinery, New York, NY, USA, 2023, p. 533–544. doi:10.1145/3543507.3583464.

[32] J. Saramäki, M. Kivelä, J.-P. Onnela, K. Kaski, J. Kertész, Generalizations of the clustering coefficient to weighted complex networks, Phys. Rev. E 75 (2007) 027105. doi:10.1103/PhysRevE.75.027105.

[33] L. Page, S. Brin, R. Motwani, T. Winograd, The pagerank citation ranking: Bringing order to the web, World Wide Web Internet And Web Information Systems 54 (1998). doi:10.1.1.31.1768.

[34] M. Mcpherson, L. Smith-Lovin, J. Cook, Birds of a feather: Homophily in social networks, Annu Rev Sociol 27 (2001) 415–444. doi:10.1146/annurev.soc.27.1.415.

[35] P. Sen, G. M. Namata, M. Bilgic, L. Getoor, B. Gallagher, T. Eliassi-Rad, Collective classification in network data, AI Magazine 29 (2008). doi:10.1609/aimag.v29i3.2157.

[36] O. Shchur, M. Mumme, A. Bojchevski, S. Günnemann, Pitfalls of graph neural network evaluation, CoRR abs/1811.05868 (2018).

[37] H. Pei, B. Wei, K. C. C. Chang, Y. Lei, B. Yang, Geom-gcn: Geometric graph convolutional networks, in: 8th International Conference on Learning Representations, ICLR 2020, 2020. doi:10.48550/arXiv.2002.05287.

[38] J. Tang, J. Sun, C. Wang, Z. Yang, Social influence analysis in large-scale networks, in: Proceedings of the ACM SIGKDD International Conference on Knowledge Discovery and Data Mining, 2009, pp. 807–816. doi:10.1145/1557019.1557108.

[39] A. R. Benson, R. Abebe, M. T. Schaub, A. Jadbabaie, J. Kleinberg, Simplicial closure and higher-order link prediction, Proceedings of the National Academy of Sciences of the United States of America 115 (2018). doi:10.1073/pnas.1800683115.

[40] B. Rozemberczki, C. Allen, R. Sarkar, Multi-scale attributed node embedding, Journal of Complex Networks 9 (2021). doi:10.1093/comnet/cnab014.

[41] P. Veličković, A. Casanova, P. Liò, G. Cucurull, A. Romero, Y. Bengio, Graph attention networks, in: 6th International Conference on Learning Representations, ICLR 2018 - Conference Track Proceedings, 2018, pp. 10–48550. doi:10.1007/978-3-031-01587-8_7.

[42] S. Luan, C. Hua, Q. Lu, J. Zhu, M. Zhao, S. Zhang, X. Chang, D. Precup, Is heterophily A real nightmare for graph neural networks to do node classification?, CoRR abs/2109.05641 (2021). arXiv:2109.05641.

[43] S. Brody, U. Alon, E. Yahav, How attentive are graph attention networks?, in: ICLR 2022 - 10th International Conference on Learning Representations, 2022. doi:10.48550/arXiv.2105.14491.

[44] M. Liu, Z. Wang, S. Ji, Non-local graph neural networks, IEEE Transactions on Pattern Analysis and Machine Intelligence 44 (2022). doi:10.1109/TPAMI.2021.3134200.

[45] M. Zou, Z. Gan, Y. Wang, J. Zhang, D. Sui, C. Guan, S. Leng, Unig-encoder: A universal feature encoder for graph and hypergraph node classification, Pattern Recognition 147 (2024). doi:10.1016/j.patcog.2023.110115.

[46] Y. Shi, Z. Huang, S. Feng, H. Zhong, W. Wang, Y. Sun, Masked label prediction: Unified message passing model for semi-supervised classification, in: Z.-H. Zhou (Ed.), Proceedings of the Thirtieth International Joint Conference on Artificial Intelligence, IJCAI-21, International Joint Conferences on Artificial Intelligence Organization, 2021, pp. 1548–1554, main Track. doi:10.24963/ijcai.2021/214.

[47] J. Chen, C. Liu, K. Gao, G. Li, K. He, Tokenized graph transformer with neighborhood augmentation for node classification in large graphs, arXiv (2023). doi:10.48550/arXiv.2206.04910.

[48] H. Yang, J. Fang, J. Wu, D. Li, Y. Wang, Z. Zheng, Soft label enhanced graph neural network under heterophily, Knowledge-Based Systems 309 (2025) 112861. doi:https://doi.org/10.1016/j.knosys.2024.112861.

[49] M. Fey, J. E. Lenssen, Fast graph representation learning with PyTorch Geometric, in: ICLR Workshop on Representation Learning on Graphs and Manifolds, 2019.

[50] A. Paszke, S. Gross, F. Massa, A. Lerer, J. Bradbury, G. Chanan, T. Killeen, Z. Lin, N. Gimelshein, L. Antiga, A. Desmaison, A. Köpf, E. Yang, Z. DeVito, M. Raison, A. Tejani, S. Chilamkurthy, B. Steiner, L. Fang, S. Chintala, Pytorch: An imperative style, high-performance deep learning library (12 2019). doi:10.48550/arXiv.1912.01703.